\documentclass{jfm_modified_2016}
\usepackage{graphicx, caption, subfig}
\usepackage{amsmath, amssymb}
\usepackage[normalem]{ulem}
\usepackage{mathtools}
\usepackage{xspace}

\usepackage{enumitem}

\pdfmapfile{+rsfso.map}
\DeclareSymbolFont{rsfso}{U}{rsfso}{m}{n}
\DeclareSymbolFontAlphabet{\mathscr}{rsfso}

\usepackage{bm}
\usepackage{microtype}

\usepackage[pdfauthor={SJ and PHT}]{hyperref}
\usepackage[usenames, dvipsnames]{xcolor}
\hypersetup{colorlinks=true,linkcolor=MidnightBlue,citecolor=MidnightBlue}

\usepackage{tikz}
\usepackage{natbib}
\usepackage{array}
\usepackage{booktabs}
\usepackage{multirow}
\usepackage{listings}
\newcommand*{\ep}{\epsilon}

\newcommand*{\im}{\mathrm{i}}
\newcommand*{\e}{\mathrm{e}}
\newcommand*{\Oh}{\mathcal{O}}

\renewcommand*{\H}{\mathscr{H}}
\newcommand*{\de}{\operatorname{d\!}{}} 
\newcommand{\dd}[2]{\frac{\de#1}{\de#2}}

\newcommand{\ddd}[2]{\frac{\de^2 #1}{\de#2^2}}
\newcommand*{\ta}{I\xspace}
\newcommand*{\tb}{II\xspace}
\newcommand*{\tc}{III\xspace}
\newcommand*{\Fr}{\mathrm{F}}
\newcommand*{\T}{\mathrm{T}}

\newcommand*{\ku}{k_\text{up}}

\newcommand*{\Td}{\T_\text{down}}
\newcommand*{\kd}{k_\text{down}}

\newcommand*{\hu}{h_\text{up}}
\newcommand*{\hd}{h_\text{down}}
\newcommand*{\Uu}{U_\text{up}}
\newcommand*{\Ud}{U_\text{down}}
\newcommand*{\Udu}{U_\text{d/u}}
\newcommand*{\hdu}{h_\text{d/u}}

\usepackage{mathabx}
\usepackage{esint} 
\def\Xint#1{\mathchoice
   {\XXint\displaystyle\textstyle{#1}}%
   {\XXint\textstyle\scriptstyle{#1}}%
   {\XXint\scriptstyle\scriptscriptstyle{#1}}%
   {\XXint\scriptscriptstyle\scriptscriptstyle{#1}}%
   \!\int}
\def\XXint#1#2#3{{\setbox0=\hbox{$#1{#2#3}{\int}$}
     \vcenter{\hbox{$#2#3$}}\kern-.5\wd0}}

\def\YYint#1#2#3{{\setbox0=\hbox{$#1{#2#3}{\int}$}
     \vcenter{\hbox{\scalebox{1}[-1]{$#2#3$}}}\kern-.5\wd0}}

\def\dashint{\Xint-}

\title[Gravity-capillary waves in reduced models]{Gravity-capillary waves in reduced models for wave-structure interactions}

\author{Sean Jamshidi\aff{1} 
 \and Philippe H. Trinh\aff{2}\corresp{\email{p.trinh@bath.ac.uk}}}

\affiliation{
\aff{1}Department of Mathematics, University College London, London WC1E 6BT, UK
\aff{2} Department of Mathematical Sciences, University of Bath, Bath BA2 7AY, UK
}

\pubyear{XXXX}
\volume{YYY}
\pagerange{xxx--xxx}
\date{\today~[Draft]}

\begin{document}

\maketitle

\begin{abstract}


\noindent This paper is concerned with steady-state subcritical gravity-capillary waves that are produced by potential flow past a wave-making body. Such flows are characterised by two non-dimensional parameters: the Froude number, $\Fr$, and the inverse-Bond number, $\T$. When the size of the wave-making body is formally small there are two qualitatively different flow regimes and thus a single bifurcation curve in the $(\Fr, \T)$-plane. If, however, the size of the obstruction is order one then, in the limit $\Fr, \, \T  \to 0$, \cite{Trinh2013} have shown that the bifurcation curve widens into a band, within which there are four new flow regimes [J. Fluid Mech. 724, pp.~392--424]. Here, we use results from exponential asymptotics to show how, in this low-speed limit, the water-wave equations can be asymptotically reduced to a single differential equation, which we solve numerically to confirm one of the new classes of waves. The issue of numerically solving the full set of gravity-capillary equations for potential flow is also discussed.

\end{abstract}

\section{Introduction} \label{sec:intro}

\begin{figure}
\includegraphics[width=1.0\textwidth]{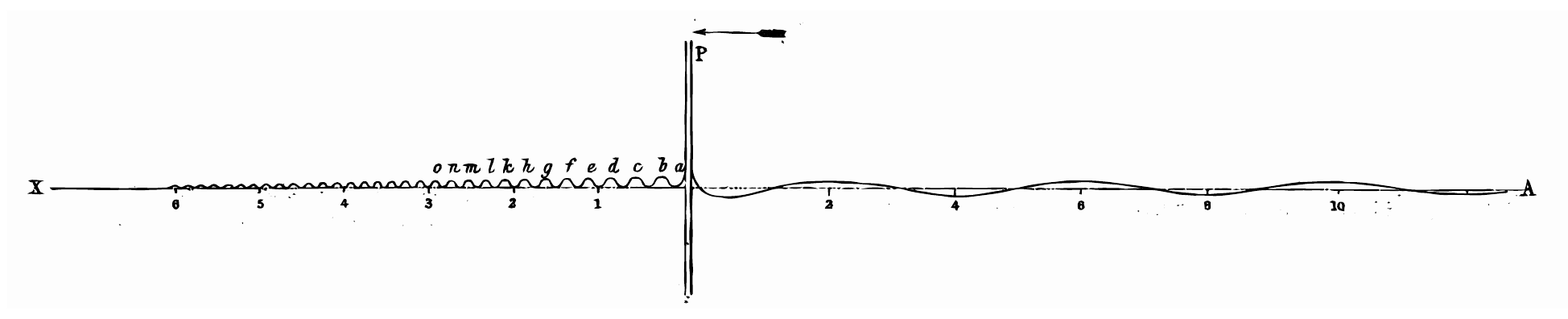}
\caption{Original illustration from Plate 57 of \cite{russell_1845_report_on} depicting the short-wavelength capillary ripples upstream and long-wavelength gravity waves downstream of a fishing line. Note that in Russell's illustration, viscous effects are included. \label{fig:russell}}
\end{figure}

\noindent The subject of this paper is the accurate calculation of steady-state free-surface flows where a disturbance has caused the emergence of distinct groups of gravity and capillary waves. In particular, we will discuss flow regimes that exist in the low-speed limit and numerically confirm the existence of a new class of waves predicted by \cite{Trinh2013}. 

The canonical application of free-surface flow is the fishing-line problem first reported by \cite{russell_1845_report_on}. Figure \ref{fig:russell} corresponds to Russell's original illustration of a two-dimensional cross-section of the free-surface for uniform flow past a fishing line. Inspired by Russell's report, \cite{rayleigh1883form} modelled the fishing line as a point pressure force and demonstrated that waves of constant amplitude exist when the flow is above a certain speed (approximately $23 \text{ cm/s}$ in water of infinite depth). The disturbances upstream of the fishing line are capillary waves, principally governed by surface tension, while those downstream are due to gravity (see also \S~{271} of \citealt{lamb_book}).

Then, in the latter half of the 20th century, it became possible to compute numerical solutions of the full nonlinear potential-flow equations for steady-state problems involving different geometries. As it pertains to the gravity-only problem, we mention as examples the works of \cite{vanden-broeck_1977_computation_of,king1987free,king1990free,forbes1982free}, although interest continues well into recent years in \emph{e.g.} \cite{puaruau2002nonlinear,binder2007effect,binder_2013_free-surface_flow}. For reasons that will become clear, the numerical study of the types of steady-state gravity-capillary profiles illustrated in figure \ref{fig:russell} is a much more difficult problem. 

Like the methodology developed in \cite{rayleigh1883form}, theoretical analysis of such wave-structure free-surface problems usually begins by linearising the governing equations about a small parameter, say $\delta$, related to the size of the obstruction (linear geometrical theory). In addition to $\delta$, gravity-capillary flows are typically characterized by two non-dimensional parameters: the Froude number, $\Fr$ and the inverse-Bond number, $\T$, defined by
\begin{equation} \label{eq:FT}
\Fr = \frac{U}{\sqrt{gL}} \qquad \text{and} \qquad \T = \frac{\sigma}{\rho g L^2}.
\end{equation}
Here, $U$ and $L$ are the chosen velocity and length scales, $\rho$ is density and $\sigma$ is the surface tension parameter. As is convention in water-wave studies, we refer to $\T$ directly as the Bond number. In this work, we focus on subcritical flows, \emph{i.e.} those with $\Fr < 1$. 

As explained by \cite{forbes1983free}, linear geometrical theory indicates that for $\delta \ll 1$ there exists a critical curve, $\T = \T_G(\Fr)$, that divides the subcritical part of the $(\Fr, \T)$ plane into two regions. Solutions with $\T < \T_G(\Fr)$ are called Type \ta and are associated with the constant-amplitude waves of figure \ref{fig:russell}, while solutions with $\T > \T_G(\Fr)$ are called Type \tb, and consist of localized solitary waves.

Recently, it was predicted in \cite{trinh2013_linear,Trinh2013} that several new classes of solution can occur for nonlinear geometries with $\delta = \Oh(1)$, in the limit of small Froude and Bond numbers, $\Fr, \, \T \to 0$. The authors considered the case of a right-angled step and used techniques from exponential asymptotics to show that the typical \cite{rayleigh1883form} bifurcation curve predicted for linear geometry widens into a band as the height of the step increases, and that within this band a range of new solutions can be found. This work aims to numerically confirm these new, Type \tc waves, and thus we seek to develop methods that allow us to accurately solve the governing equations for nonlinear geometries. In doing so we make use of ideas developed in \cite{trinh2016topological,Trinh2017reduced}, where computationally problematic integral terms [cf. \eqref{eq:hilbert_def}] are removed from the governing equations through a systematic asymptotic reduction in the low-speed, $\Fr, \, \T \to 0$ limit.

\subsection{The radiation problem} \label{sec:radcon}
Numerical solutions of steady-state free-surface problems require the use of effective radiation conditions ---that no energy may come from infinity--- on the edges of a truncated computational domain. However it is often unclear exactly how one can impose such a condition in practice, particularly when capillary waves are present. \cite{stoker_book} had previously provided several remarks on this unexpectedly challenging aspect of determining steady-state flows. As he writes (p.~175):  
\begin{quotation}
\noindent \emph{\ldots it is by no means clear a priori what conditions should be imposed at infinity in order to ensure the uniqueness of a simple harmonic solution [\ldots] the steady state problem is unnatural---in the author's view, at least---because a hypothesis is made about the motion that holds for all time, while Newtonian mechanics is basically concerned with the prediction---in a unique way, furthermore---of the motion of a mechanical system from given initial conditions.} 
\end{quotation}
As Stoker notes, one should in principle formulate and solve an initial-value problem where it is often sufficient to only impose boundedness of the solution at infinity. Afterwards, the time-dependent solution may then be evolved to a steady state. In practice, however, there are a myriad of reasons why solving a steady-state formulation may be necessary---beyond simple computational efficiency. For instance, the study of solutions of the direct steady-state model may yield a wealth of information about the physical or mathematical structure of the problem; information that cannot be easily obtained through time-dependent experiments.

Apart from specialized configurations (\emph{e.g.} localized solitary waves), there has not yet been a proposed solution of the radiation problem, particularly for the case of Type \ta solutions. One approach, suggested by \cite{grandison2006truncation} approximates the solution outside of the main computational domain using a Fourier series whose coefficients are found as part of the solution. Although this method was able to successfully compute a number of Type \ta solutions for flow over a small circular cylinder (non-dimensional radius $\delta = 0.05$), we have found that convergence is difficult to obtain for nonlinear geometries. Other options for correctly enforcing the radiation condition include the addition of artificial viscosity, say $\mu$ [as in \citealt{puaruau2007three}] or by solving the full time-dependent problem [as in \cite{puaruau2010time,moreira2010nonlinear}]. However, in both cases, it is not clear that taking the limit $\mu \to 0$ or $t \to \infty$ will allow recovery of the full set of solutions when $\mu = 0$ and $t = \infty$. 

Our main strategy forwards is to explain how, in the limit $\Fr, \T \to 0$, the governing equations may be reduced to a single second-order differential equation, for which it is clear how the radiation condition should be applied. We solve this reduced model as a boundary-value problem, and use it to verify the new waves predicted in \cite{Trinh2013}. The reduced model is derived in sections \ref{sec:mathfo}--\ref{sec:reduced}, and then solved numerically in section \ref{sec:results}. Sections \ref{sec:disprel}-\ref{sec:smallep} give a simple explanation for one of the new classes of waves.

\section{An explanation of the new waves via the dispersion relations} \label{sec:disprel}

\noindent Here, we give a simple explanation of one of the new classes of gravity-capillary waves that were previously described using the more sophisticated exponential asymptotics. This is done by comparing upstream and downstream dispersion relations, and is similar in spirit to arguments in \cite{binder2007effect}. 

\subsection{The upstream dispersion relation} \label{sec:updisp}

\begin{figure} \centering
\includegraphics{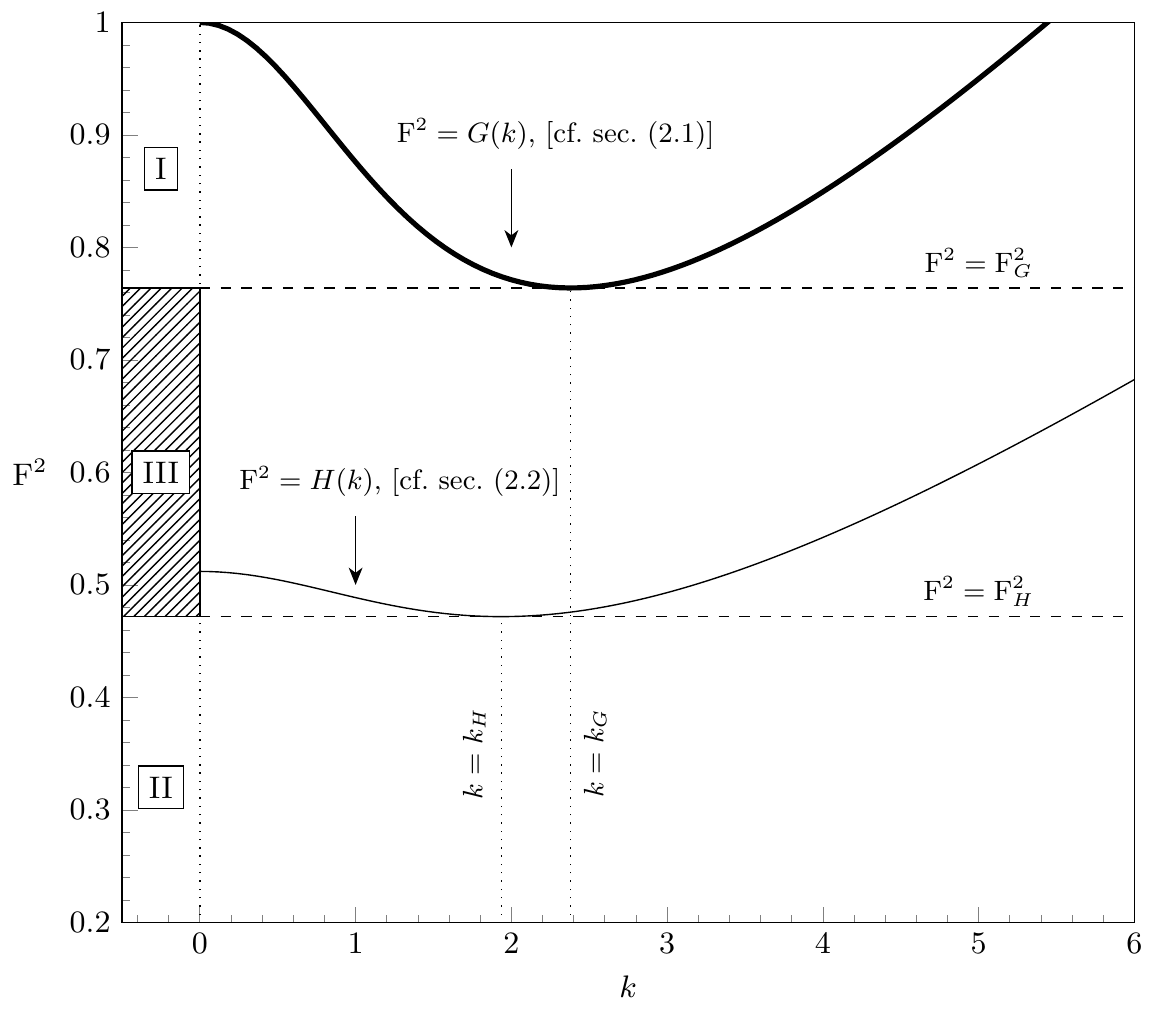}
\caption{Upstream (thick) and downstream (thin) dispersion relations. The upstream relation $G(k)$ is given by \eqref{Fup}, and the downstream dispersion relation $H(k)$ is given by \eqref{Fup2}. Both are plotted with $T = 0.15$, and the depth ratio in $H(k)$ is $\hdu = 0.8$. Constant-amplitude waves exist for values of $\Fr$ that intersect the dispersion curve. Thus there is a region, marked `\tc', where constant-amplitude waves only exist downstream. \label{fig:kF2_smallT}}
\end{figure}

The dispersion relation that governs linear perturbations of wavenumber $\tilde{k}$ from uniform flow of speed $U$ and depth $h$ is [cf. (2.90) in \cite{vanden2010gravity}]
\begin{equation} \label{nondimdisp}
  U^2 = \left[ \frac{g}{\tilde{k}} + \frac{\sigma}{\rho} \tilde{k} \right] \tanh (\tilde{k}h).
\end{equation}
In most typical formulations, the dispersion relation is defined in the context of the upstream quantities, thus re-scaling the expression above, we have
\begin{equation} \label{Fup}
  \Fr^2 = \left[ \frac{1}{k} + \T k \right] \tanh (k) \equiv G(k; \T).
\end{equation}
In \eqref{Fup}, we have used the mean upstream channel height, $L = \hu$, and velocity, $U = \Uu$, for the definitions of the Froude and Bond numbers [cf. \eqref{eq:FT}]. The wavenumber has been non-dimensionalized via 
\begin{equation}
k = \tilde{k} \hu.
\end{equation}

Thus, for a given value of $\T$, the dispersion relation $\Fr^2 = G(k)$ can be plotted in the $(k, \Fr^2)$-plane. An example profile of $G(k)$, with $T = 0.15$, is shown in figure \ref{fig:kF2_smallT}. 

Note that if $\T < 1/3$, the upstream dispersion relation $G(k)$ has a minimum at $(k_G, \Fr_G ^2)$. The existence of the local minimum leads to two possible types of solutions within subcritical flows ($\Fr < 1$):
 
\begin{enumerate}[label={Type \Roman*:}, leftmargin=*, align = left, labelsep=0.4cm, topsep=3pt, itemsep=4pt, itemindent=0pt ]
\item If the Froude number lies within the band $\Fr_G < \Fr < 1$, labeled \ta in the figure, then there exists two distinct real wave numbers. The smaller of the two wavenumbers corresponds to gravity waves and the larger to capillary waves. The solution consists of constant-amplitude capillary waves upstream and constant-amplitude gravity waves downstream. 

\item If, however, the Froude number is less than the minimum of $G(k)$ and in regions \tb and \tc of figure \ref{fig:kF2_smallT}, \emph{i.e.} $0 < \Fr < \Fr_G$, then solutions to \eqref{Fup} are complex-valued. Thus the wave-trains decay in the far field. 
\end{enumerate}

\subsection{The downstream dispersion relation}


In addition to the two classes described above, the analysis of \cite{Trinh2013} uses techniques from exponential asymptotics to predict several new classes of waves that occur in the low-Froude, low-Bond limit for finite-bodied (nonlinear) objects. However there is a simple explanation of one of these new classes. The key idea is that if the size of the step is sufficiently large, then the velocity and length scales downstream of the step must be re-defined. Consequently, a different dispersion relation applies downstream. 
 
We introduce the height and velocity ratios, 
\begin{equation}
  \hdu = \frac{\hd}{\hu} \quad \text{and} \quad \Udu = \frac{\Ud}{\Uu},
\end{equation}
and the downstream Froude and Bond numbers, as well as downstream wavenumbers,
\begin{equation} \label{ratios}
  \Fr =  \Fr_\text{down}\frac{\sqrt{\hdu}}{\Udu}, \qquad 
\T = \Td \hdu^2, \qquad k = \frac{\kd}{\hdu},
\end{equation}
where by mass conservation $\hdu = 1/\Udu$.

Applying \eqref{nondimdisp} to these downstream quantities, and re-writing in terms of upstream quantities, we must have 
\begin{equation} \label{Fup2}
\Fr^2 = \hdu^2 \left[ \frac{1}{k} + \T k \right] \tanh (\hdu k) 
\equiv H(k; \T, \hdu).
 \end{equation} 
Thus in addition to the upstream dispersion relation $G(k)$, we plot the downstream dispersion relation $H(k)$ on figure \ref{fig:kF2_smallT}, for the same value of $\T$ and with $\hdu = 0.8$. The downstream dispersion curve $H(k)$ has a local minimum at $(k_H, \Fr_H^2)$. Consequently, if the upstream Froude number is selected from within the band $\Fr_H < \Fr < \Fr_G$ then, although the upstream dispersion relation predicts decaying waves, the downstream dispersion relation indicates the presence of constant-amplitude waves. The result is a third class of solutions, in addition to those described in section~\ref{sec:updisp}. 
\begin{enumerate}[label={Type \Roman*:}, leftmargin=*, align = left, labelsep=0.4cm, topsep=3pt, itemsep=4pt, itemindent=0pt ]
\setcounter{enumi}{2}
\item Decaying waves upstream and constant-amplitude gravity waves downstream. This regime is labeled as \tc in figure \ref{fig:kF2_smallT}.
\end{enumerate}

\subsection{A study of the $(\Fr, \T)$-plane} \label{sec:ft}
\begin{figure} \centering
\includegraphics[width=0.9\textwidth]{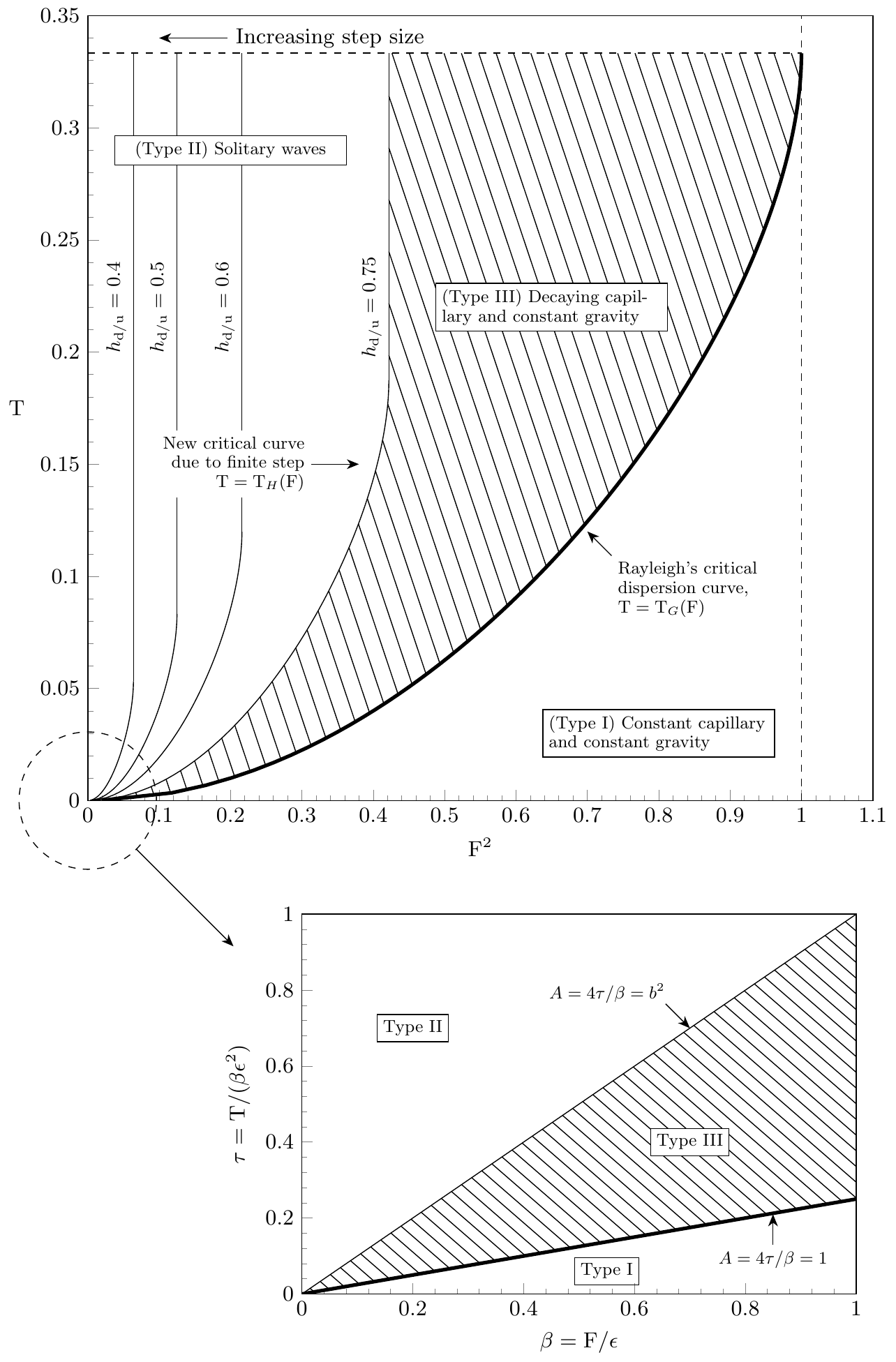}
\caption{(Upper) Upstream (thick) and downstream (thin) bifurcation curves in the $(\Fr,\T)$ plane, as discussed in section \ref{sec:ft}. For the case of $\hdu = 0.75$, the shaded area between them gives the region where new waves are possible. As $\hdu$ decreases, the step size becomes larger and the area between the two bifurcation curves grows. (Lower) The different regions in the low-speed limit, where the boundaries of the bifurcation curve are described in terms of the step-height parameter $b$ [cf. section \ref{sec:smallep}]. \label{fig:FTbif}}
\end{figure}

The bifurcation between constant-amplitude waves and decaying waves can also be viewed in the $(\Fr, \T)$ plane, and this is shown in the top portion of figure~\ref{fig:FTbif}. There are two important features.

Firstly, notice the thick line in the figure. This is essentially Rayleigh's critical dispersion curve discussed in sections~\ref{sec:intro} and \ref{sec:updisp}. That is, for $\Fr < 1$ and $\T < 1/3$, there exists a critical curve $\T = \T_G(\Fr)$, which corresponds to the minimum over $k$ of the upstream dispersion relation $G(k; \T)$ for each $\T$, where $G(k)$ is given in \eqref{Fup}. For linear geometries, this is the only critical curve. Configurations below the curve have constant-amplitude waves (Type \ta solutions), and configurations above the curve have decaying waves (Type \tb solutions). 

Secondly, notice the sequence of thinner curves in the figure. For the nonlinear step a second critical curve $\T = \T_H(\Fr)$ exists, which corresponds to minima in the downstream dispersion relation $H(k; \T)$, given by \eqref{Fup2}. New solutions (Type \tc) with decaying capillary waves and constant-amplitude gravity waves are located in the region of the $(\Fr, \T)$ plane that lies between these two curves. Note the area of this region increases with the size of the step. The thin lines in figure \ref{fig:FTbif} show the downstream critical curve $\T_H(\Fr)$ for various values of the depth ratio $\hdu$, with the shaded area giving the region where new waves can be found for the particular case of $\hdu = 0.75$. 

\section{Asymptotics in the small Froude-Bond limit} \label{sec:smallep}

\noindent Although our study of linear dispersion relations in the previous sections has given an insight into the new regimes in the $(\Fr, \T)$-plane, it is important to recall that \eqref{Fup} and \eqref{Fup2} are derived from the study of the free-surface problem assuming a flat-bottomed channel, which is only valid sufficiently far upstream and downstream. To derive a spatially-dependent dispersion relation that connects the upstream and downstream regions, it is necessary to use the more complex exponential asymptotics, which are valid as $\Fr, \T \to 0$. Since it is in this low-speed limit where the new waves are predicted, we re-write the dispersion relation \eqref{nondimdisp} with 
\begin{equation}
\Fr^2 = \beta \epsilon \qquad \text{and} \qquad \T = \beta \tau \epsilon^2,
\end{equation}
for $\ep \ll 1$, following scalings from previous works \citep{trinh2013_linear, Trinh2013}.

In the limit $\epsilon \to 0$, the wavenumber scales like $k = \hat{k}/\epsilon$. It will be shown below in section \ref{sec:mathfo} that, to leading-order in $\ep$, the asymptotic values of the speed downstream and upstream are $\sqrt{b}$ and unity respectively, where $b$ is a parameter describing the step height. The velocity scale ratio for flow over a step is therefore
\begin{equation}
\Udu = \sqrt{b} + \Oh(\epsilon) \sim \frac{1}{\hdu}.
\end{equation}

In this low-speed regime, $\tanh (k) \to 1$ and the dispersion relation \eqref{Fup} becomes
\begin{equation}
\beta \tau \ep^2 k^2 - \beta \ep k + 1 = 0 \label{eq:smallepdisp}
\end{equation}
with solutions 
\begin{equation} \label{eq:kup}
  \ku = \frac{1 \pm \sqrt{1 - A}}{2\tau \ep},
\end{equation}
where $A = 4\tau/\beta$. This is the typical small-step wavenumber in the limit of small Froude/Bond numbers, and predicts a band $A \in [0, 1]$ where wavenumbers are real.

However, for a large step we must use \eqref{Fup2} and then \eqref{ratios} to calculate the downstream wavenumbers as
\begin{equation} \label{eq:kdown}
  \kd = \ku \hdu = \frac{b \pm \sqrt{b^2 - A}}{2\tau \sqrt{b} \ep}.
\end{equation}
Thus, if the Froude and Bond numbers are chosen so that $A \in [1,b^2]$ then an additional regime is predicted, before the localised solitary waves emerge for $A>b^2$. The three different regions are illustrated in the lower portion of figure \ref{fig:FTbif}. The function $\chi(\phi)$ that gives the spatially-varying phase of the waves is found in section 3 of \cite{Trinh2013}. Comparison of (3.9) in that work with the expressions above shows that the wavenumbers here are just the far-field limits of $\chi$, as $|\phi| \to \infty$. 

In both \eqref{eq:kup} and \eqref{eq:kdown}, the plus root becomes singular as $\tau \to 0$. This root therefore corresponds to capillary waves, and the minus root must correspond to gravity waves. This removes any ambiguity introduced by the extra dispersion relation, as we have exactly one upstream wavenumber that satisfies the radiation condition. The wavenumbers $\ku$ and $\kd$ can thus be used to verify that numerical solutions satisfy the radiation condition. However these wavenumbers can also be used to \emph{prescribe} the behaviour of the flow at infinity, and indeed this approach will later allow us to numerically confirm the existence of the new waves in the reduced model. First, we expand on the discussion of section \ref{sec:radcon}, and illustrate the issues surrounding the numerical solution of the full problem. 

\section{Boundary integral formulation} \label{sec:mathfo}

\begin{figure}
\includegraphics{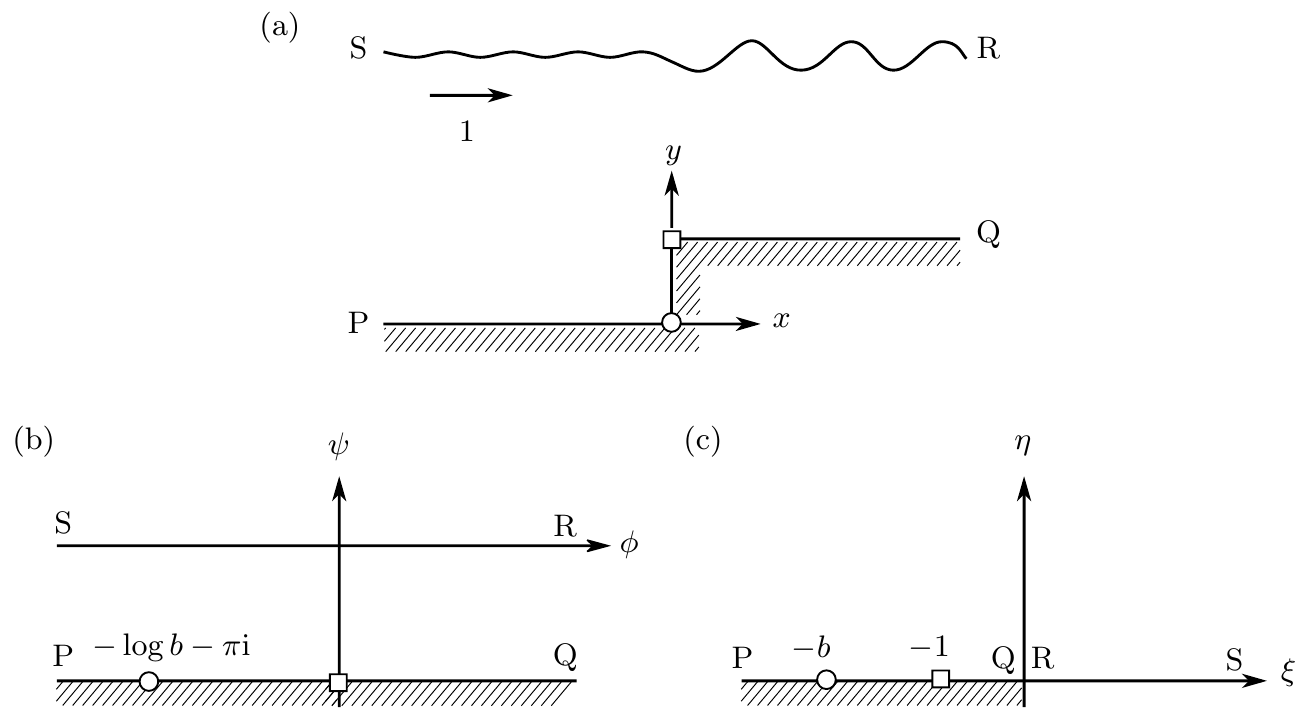}
\caption{Flow over a step in a channel shown in the (a) physical $z$-plane, (b) potential $w$-plane, (c) upper half-$\zeta$-plane. In our non-dimensionalization, the step corner (square) is chosen to be $\zeta = -1$ and the stagnation point (circle) is chosen to be $\zeta = -b$ with $b > 1$. The labels PQRS show the orientation of the different mappings with respect to upstream and downstream locations. \label{fig:planes}}
\end{figure}

\noindent Consider steady, two-dimensional potential flow of an incompressible fluid in a channel with upstream velocity $U$, and a prescribed length scale $L$. The physical plane is shown in figure \ref{fig:planes}(a). The flow is non-dimensionalised using these characteristic scales, and we introduce complex physical coordinates, $z = x + \im y$, with $x$ pointing along the channel, as well as a complex potential $w = \phi + \im \psi$. The governing equations for the fluid are then 
\begin{subequations} \label{eq:origsys}
\begin{align}
\text{Laplace's equation}&:& \nabla^2 \phi &= 0 && \text{inside the fluid}, \label{eq:Laplace} \\
\text{Kinematic condition}&:& \nabla \phi \cdot \bm{n} &= 0 && \text{on solid/free surfaces}, \\
\text{Bernoulli's equation}&:& \Fr^2 |\nabla \phi|^2/2 + y + \T \kappa &= \text{const.} && \text{on the free surface}. \label{eq:bernoulli_dim}
\end{align}
\end{subequations}
In the kinematic condition, $\bm{n}$ denotes a normal of the channel boundary and, in Bernoulli's equation, the curvature, $\kappa$, is defined to be positive when the centre of curvature lies inside the fluid. The derivation of this standard governing formulation is given in \emph{e.g.} Chap.~2 of \cite{vanden2010gravity}. 

In the analysis of the potential plane, it is convenient to choose the length scale to be
\begin{equation}
  L = \frac{h_\text{up}}{\pi},
\end{equation}
where $h_\text{up}$ is the upstream height of the channel, and thus the non-dimensional height is set to $\pi$. We also non-dimensionalise the velocity according to its upstream value $U$, so that the non-dimensional flux is $\pi$. Setting $\psi = 0$ on the free surface, it must then follow that $\psi = - \pi$ on the channel bottom and consequently, the flow lies within a strip PQRS in the $w$-plane shown in figure~\ref{fig:planes}(b).   

It is convenient to further map the strip-flow in the $w$-plane to the upper half-$\zeta$-plane using the conformal map
\begin{equation} \label{eq:zeta}
  \zeta = \xi + \im \eta = \e^{-w} = \e^{-(\phi + \im \psi)}.
\end{equation}
Thus, the free surface and channel bottom are both mapped to $\eta = 0$, with $\xi > 0$ the free surface and $\xi < 0$ the channel bottom. The flow in the upper half-$\zeta$-plane is shown in figure~\ref{fig:planes}(c), with the solid and fluid boundaries, marked PQRS, lying along the real axis.  

Next, we introduce the complex velocity, $\dd{w}{z}$ by writing
\begin{equation} \label{eq:dwdz}
  \dd{w}{z} = q \e^{-\im \theta},
\end{equation}
for fluid speed $q$ and streamline angle, $\theta$. Below, we shall recast the original system \eqref{eq:origsys} in terms of $q$ and $\theta$, either written as functions of $w$ or $\zeta$.

Following the standard procedure, the requirement that $\phi$ satisfies Laplace's equation \eqref{eq:Laplace} is equivalently re-stated as a boundary-integral relationship between $q$ and $\theta$ applied on the fluid and solid surfaces. For points on the free-surface, $\xi \geq 0$, this allows us to write [cf. eqn (3.4) in \citealt{trinh2013_linear}]
\begin{equation} \label{eq:logqtmp}
  \log q(\xi) = -\frac{1}{\pi}\left[ \int_{-\infty}^0
\frac{\theta(\xi')}{\xi'-\xi} \ \de{\xi'} + \dashint_{0}^{\infty} \frac{\theta(\xi')}{\xi'-\xi} \ \de{\xi'}\right].
\end{equation}
Above, the first integral on the right hand-side is known (since we assume that the channel geometry is specified via $\theta$ for $\xi \leq 0$). The second integral is the Hilbert transform applied to the free surface, and the dash across the integral sign indicates a principal value.

In this work, we consider for convenience the case of flow over a right-angled step, given by specifying the streamline angles,
\begin{equation} \label{eq:thetastep}
  \theta(\xi) = \begin{cases}
  0 & -\infty < \xi < -b, \\
  \pi/2 & -b < \xi < -1, \\ 
  0 & -1 < \xi < 0,
  \end{cases}
\end{equation}
where $b > 1$ and $\xi = -b$ is the image of the stagnation point of the step in the $\zeta$-plane. Substitution of \eqref{eq:thetastep} into the first integral on the right hand-side of \eqref{eq:logqtmp} and integrating yields 
\begin{equation} \label{eq:leadingspeed}
-\frac{1}{\pi} \int_{-\infty}^0
\frac{\theta(\xi')}{\xi'-\xi} \ \de{\xi'} = \log q_s \quad \text{where} \quad q_s(\xi) \equiv \left(\frac{\xi + b}{\xi + 1}\right)^{1/2}.
\end{equation}
In \eqref{eq:leadingspeed}, we have defined the important \emph{`shape-function'}, $q_s$, which encodes the channel-bottom geometry and plays a critical role in the work that follows. 

Returning to the second integral on the right hand-side of \eqref{eq:logqtmp}, we shall also define the Hilbert transform operator $\H$, defined on the free surface $\xi \geq 0$ by
 \begin{equation}
 \H [\theta](\xi) \equiv  -\frac{1}{\pi}\dashint_{0}^\infty
\frac{\theta(\xi')}{\xi'-\xi} \, \de{\xi'}, \label{eq:hilbert_def}
 \end{equation}
 so that the boundary integral equation \eqref{eq:logqtmp}  can be written compactly as 
\begin{subequations} \label{eq:gcwave_re} 
 \begin{equation} \label{eq:gcwave_bdint_re} 
\log q(\xi)= \log q_s(\xi) + \H[\theta](\xi), 
 \end{equation}
 which is to be satisfied along the free surface, $\xi \geq 0$. 

To close the system, Bernoulli's equation in \eqref{eq:bernoulli_dim} is written in terms of $q$ and $\theta$ using \eqref{eq:dwdz}. Thus, along the free surface $\psi = 0$, we have
\begin{equation} \label{eq:gcwave_dyn_re}
\Fr^2 \left[ q^2 \dd{q}{\phi} \right]
- \T \left[ q^2 \ddd{\theta}{\phi} +
 q \dd{q}{\phi} \dd{\theta}{\phi} \right] =
- \sin{\theta}.
\end{equation}
\end{subequations}
The object now is to solve the integro-differential system given by the boundary integral \eqref{eq:gcwave_bdint_re} and Bernoulli's equation \eqref{eq:gcwave_dyn_re} for the unknowns, $q$ and $\theta$, along the free surface given by $\xi \geq 0$ or equivalently $-\infty < \phi < \infty$ and $\psi = 0$. Note however that for computational purposes we must write \eqref{eq:hilbert_def} as
\begin{equation} \label{eq:hilbert_num}
\H[\theta](\phi) = -\frac{1}{\pi} \left(\int_{-\infty}^{-\Phi} \
 + \dashint_{-\Phi}^{\Phi} \ 
 + \int_{\Phi}^{\infty}\right)
\frac{\theta(\e^{-\phi'})}{\e^{-\phi'}-\e^{-\phi}} (-\e^{-\phi'}) \, \mathrm{d} \phi',
 \end{equation}
where $\pm \Phi$ are the limits of the computational domain. The standard approach is to discard the first and third terms in \eqref{eq:hilbert_num}, which is known to introduce errors into the solution. These errors can be reduced by instead approximating the first and third terms by asymptotic solutions in the far-field, which was done for flow over a cylinder in \cite{grandison2006truncation}. Thus the numerical treatment of the Hilbert transform is key to satisfying the radiation condition. 
 
\subsection{Challenges in solving the full problem} \label{sec:fullprob}
\begin{figure} \centering
\subfloat[Boundary-integral solutions for flow over a step of non-dimensional height 0.2 at Froude number $\Fr^2 = 0.5$, for $\T = 0$ (dashed curve) and $T = 2.5 \times 10^{-3}$ (solid curve) with $N=800$ grid points. Shown is the surface speed $q$, as a function of the velocity potential $\phi$.]{\includegraphics[width=0.95\textwidth]{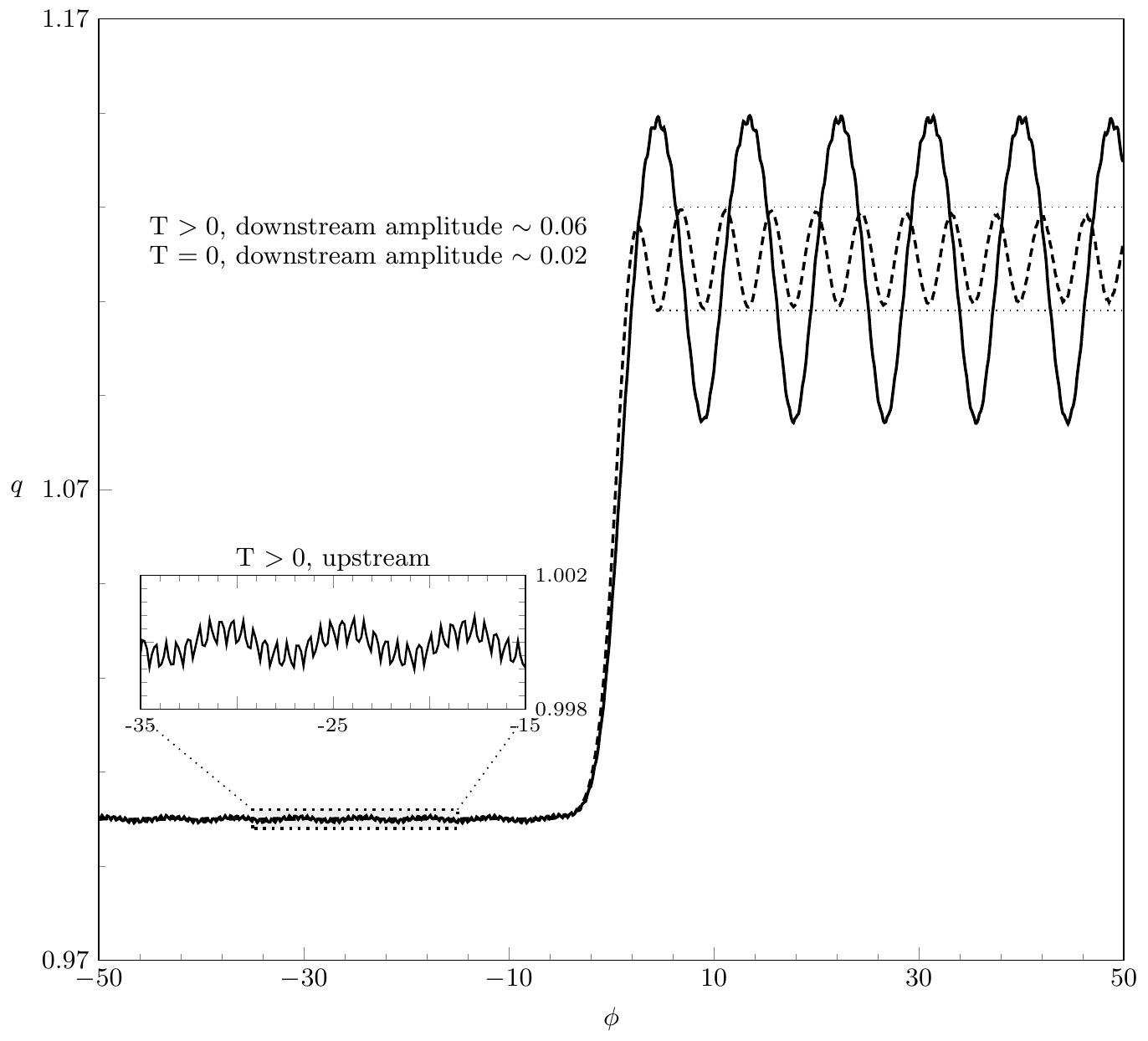}}
	
\subfloat[Fourier spectra for $q$, showing, for (i)-(iv) respectively: the upstream gravity flow, downstream gravity flow, upstream gravity-capillary flow, and downstream gravity-capillary flow. On each spectrum, `G' denotes the gravity wavenumber predicted by the dispersion relation \eqref{eq:smallepdisp}. The predicted capillary wavenumber for the $T > 0$ flow is $k = 198$.]{\includegraphics[width=0.95\textwidth]{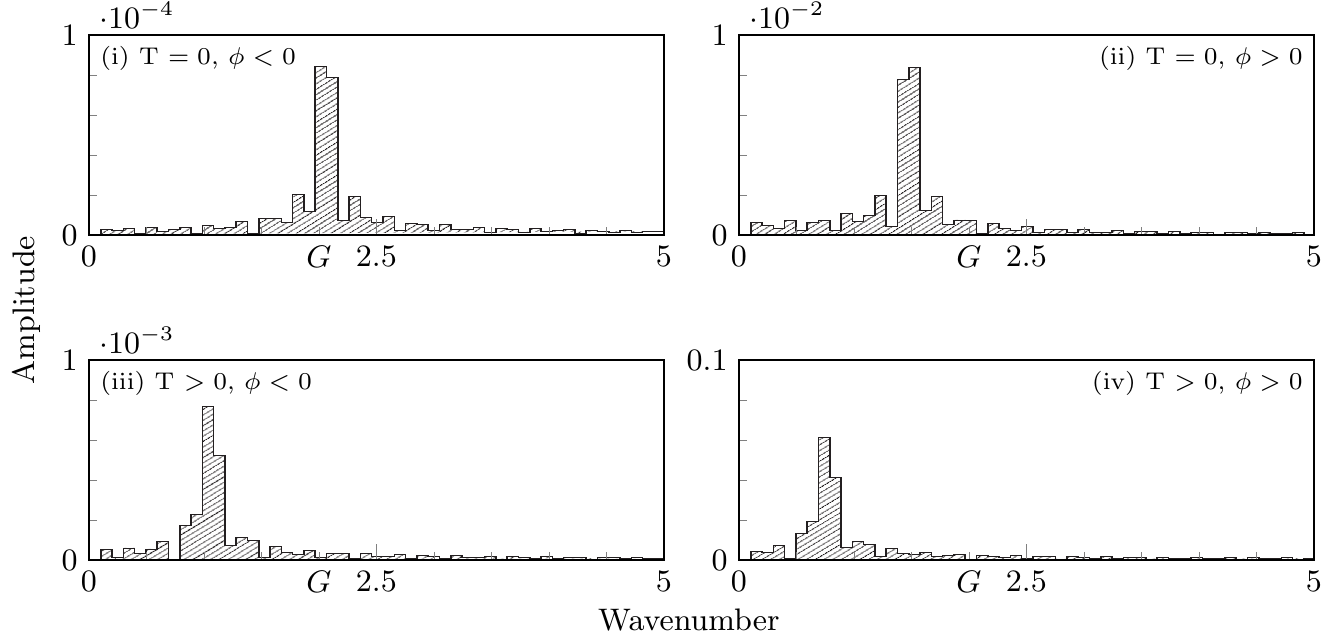}}
\caption{Solutions to the boundary-integral equations \eqref{eq:gcwave_re}. \label{fig:stepforbes}}
\end{figure}

The main challenge that confronts us in solving \eqref{eq:gcwave_re} is illustrated by figure \ref{fig:stepforbes}(a) which shows two numerical solutions; one without surface tension ($\T = 0$, dashed) and one with small amount of surface tension ($\T = 2.5 \times 10^{-3}$, solid). To produce figure \ref{fig:stepforbes}, we discretise equations \eqref{eq:gcwave_re} using finite differences and solve the resulting nonlinear system using Newton's method. This approach is used in the majority of the works referenced above \citep{forbes1982free, forbes1983free, king1987free} and is described in detail in \cite{vanden2010gravity}.

Figure \ref{fig:stepforbes}(b) shows the Fourier spectra of the solutions $q(\phi)$, divided into either upstream or downstream components. Spectra (i) and (ii) correspond to $\T = 0$, while spectra (iii) and (iv) correspond to $\T \neq 0$. In this and all subsequent figures, Fourier spectra are calculated as follows: first, the solution, $q$, is interpolated onto a regular grid and the relevant upstream ($\phi < 0$) or downstream portions ($\phi > 0$) extracted. The Fourier transform is taken and the results shown as a bar graph. On the (angular) wavenumber axis of each spectrum, we mark predicted gravity and capillary wavenumbers with a `G' and `C' as required, according to the predictions of sections \ref{sec:disprel} and \ref{sec:smallep}. 

There are three types of error visible in figure \ref{fig:stepforbes}.
\begin{enumerate}[label=(\roman*),leftmargin=*, align = left, labelsep=\parindent, topsep=3pt, itemsep=2pt,itemindent=0pt ]
\item The gravity-only flow (dashed curve) should consist of constant-amplitude waves downstream and a flat (or exponentially decaying) surface upstream. However, the dotted horizontal guides show that the downstream waves are distorted over the last few periods. This distortion is primarily associated with errors in evaluating the truncated Hilbert transform \eqref{eq:hilbert_def}. The transform is expressed as an integration over the whole real line, but must be truncated to a finite computational domain. 

\item Still for $\T = 0$, the Fourier spectrum (i) shows that there are small gravity waves upstream. The radiation condition requires that only capillary waves exist upstream, so these waves are unphysical.

\item Such errors become even more problematic for the flow with surface tension, $\T = 2.5 \times 10^{-3}$. The solid curve in figure \ref{fig:stepforbes}(a) demonstrates the presence of spurious waves upstream, as seen by the smaller inset. For this value of $(\Fr, \T)$ the non-dimensional capillary wavenumber is $k = 198$, but the upstream Fourier spectrum in (iii) shows that the dominant frequency is approximately one. Just as in the gravity-only flow, the solution does not satisfy the radiation condition. Furthermore, spectrum (iv) confirms that the downstream waves are not of the correct gravity-associated wavenumber, either. 
\end{enumerate}

Convergence of the numerical scheme is more difficult at larger values of $\T$. We emphasize that although the capillary waves shown in the smaller inset of figure \ref{fig:stepforbes}(a) are not well resolved, the issues (i) to (iii) are not functions of the grid spacing or numerical tolerances. 

Our figure \ref{fig:stepforbes} takes inspiration from figure 3 of \cite{forbes1983free}, who provided some initial discussion of the radiation problem for flow over a semi-circular cylinder. \cite{forbes1983free} further indicated that without knowledge of the proper boundary conditions, it was unclear how numerical solutions could be obtained for more nonlinear flows. Similar comments on the difficulty of the radiation problem appear in the works of \emph{e.g.} \cite{forbes1982free}, \cite{scullen}, \cite{puaruau2002nonlinear}, and \cite{grandison2006truncation} for a variety of free-surface problems. 

\section{Reduced models for gravity-capillary waves} \label{sec:reduced}

\noindent The preceding discussion makes clear that one problem with numerical computation of steady-state gravity-capillary waves is the truncation of the Hilbert transform, $\H$, and the inability to correctly apply the radiation condition. In fact, as was argued by \cite{trinh2016topological,Trinh2017reduced}, in the low-Froude and low-Bond limit $\epsilon \to 0$, the Hilbert transform can be systematically removed and the governing system \eqref{eq:gcwave_re} reduced. Here we present only the ideas that are most relevant to the reduction. Detailed calculations are given in the appendix. A full comparison of reduced models for gravity-only flow is given in \cite{Trinh2017reduced}.

\subsection{Removing the Hilbert transform} \label{sec:nohilb}

For the purpose of developing the reduced model, it is important to consider three main points that emerge as a result of exponential asymptotics \citep{chapman2006exponential}. Firstly, in the limit $\ep \to 0$ free-surface waves are associated with singularities in the analytic continuation of the leading-order solution $q_s$, given by \eqref{eq:leadingspeed}. These singularities lie in the complex plane, so it is necessary to write complex versions of the governing equations. That is, we extend $\phi$ to the complex plane, and re-write $\phi \mapsto w \in \mathbb{C}$.  Similarly, we re-write $\xi \mapsto \zeta \in \mathbb{C}$. Care must be taken when defining the Hilbert transform \eqref{eq:hilbert_def} in the complex plane so as to preserve the principal-value behaviour as $\zeta$ approaches the real axis. This is accomplished through a small deformation of the path of integration about the point of evaluation, which contributes a residue term. 

The complex versions of the governing equations are thus 
\begin{subequations}\label{eq:complexgov}
\begin{align}
\beta \ep q^2 q' - \beta \tau \ep^2 \left( q^2 \theta '' + q q' \theta' \right) + \sin {\theta} &= 0, \label{eq:compbern} \\
\log q - \im \theta + \hat{\H}[\theta] -\log q_s &= 0, \label{eq:comphilb}
\end{align}
\end{subequations}
where primes $(')$ denote differentiation in $w$ and where we have defined
\begin{equation}
\hat{\H}[\theta](\zeta) = \frac{1}{\pi} \int_0^{\infty} \frac{\theta (\xi')}{\xi' - \zeta} \mathrm{d} \xi'.
\end{equation}

Now in the system \eqref{eq:complexgov}, we have moved from the free surface, where $q, \, \phi \in \mathbb{R}$, to the complex upper-half plane, where $q$ and $\phi = w \in \mathbb{C}$, and so solutions are complex-valued. To return to the free surface, where the solution is real, we must account for the contribution from the lower-half plane. This is done by adding $q$ to its complex conjugate, $q^*$. Understandably, these ideas of analytic continuation are subtle, but for more details, we refer readers to \S{5.1} and \S{8.4} of \cite{Trinh2017reduced}.

Next, in the limit $\ep \to 0$, we expect that the solution, $q$, can be split into a base series $q_r$ which describes the mean flow, and a remainder term $\bar{q}$ which describes the waves [cf. \cite{ogilvie_1968_wave_resistance:,chapman2006exponential}]. The base series is the regular asymptotic expansion in powers of $\ep$ with, say, $N$ terms, and the remainder is $\Oh(\ep^N)$. Thus:
\begin{equation} \label{eq:splitting}
q = q_r + \bar{q} = \sum_{n = 0}^{N-1} \ep^n q_n + \bar{q}.
\end{equation}
We also do the same for $\theta$. The inclusion of the remainder term in \eqref{eq:splitting} is necessary because using a regular expansion alone predicts a flat surface at every algebraic order. In the low-speed limit, the waves are exponentially small and must be found with specialised techniques from exponential asymptotics.

Finally, a careful analysis of the governing equations using the split solution \eqref{eq:splitting} reveals that computation of the mean speed $q_r$ only involves the Hilbert transform of known terms, that is $\hat{\H}[\theta_r]$. Further, the remainder term $\bar{q}$, and hence the form of the waves, does not depend on the Hilbert transform of $\bar{\theta}$ at leading order. Therefore the problematic term $\hat{\H}[\bar{\theta}]$ can be ignored in an asymptotically consistent way in the limit as $\ep \to 0$. Equation \eqref{eq:comphilb} reduces to the simple relationship
\begin{equation} \label{eq:nohilb}
\bar{q} = \im q_s \bar{\theta},
\end{equation}
which can be substituted into \eqref{eq:compbern} to give an ordinary-differential equation (ODE) for $\bar{q}$, which is known as a reduced model.

\subsection{Choosing the truncation value $N$} \label{sec:truncate}
The derivation of the reduced model for an arbitrary truncation value $N$ is given in the appendix, but here we focus on the case $N = 2$. This particular value of $N$ is chosen because it captures the functional form of the waves \citep{Trinh2017reduced}. In the limit $\ep \to 0$, we expect the form of the waves to be  
\begin{equation} \label{eq:qexpansatz}
  \bar{q} \sim \mathcal{A} \mathcal{F}(w) \e^{-\chi(w)/\ep}.
\end{equation}
If the base series \eqref{eq:splitting} is truncated at $N = 1$, then the resulting $N = 1$ reduced model produces a solution with the correct phase function $\chi(w)$. The $N = 2$ reduced model produces a solution where both $\chi$ and $\mathcal{F}$ are correct. To obtain the constant pre-factor $\mathcal{A}$, one must truncate the base series at the optimal point $\mathcal{N}(\ep)$, where it is shown in \cite{chapman2006exponential} that $\mathcal{N} \to \infty$ as $\ep \to 0$. The choice of $N = 2$ therefore correctly predicts the form of the waves, up to a multiplicative constant $\mathcal{A}$ \citep{Trinh2017reduced}.  

\subsection{The $N = 2$ reduced model for the low-Froude low-Bond limit}
The first two terms in the truncated base series \eqref{eq:splitting} are:
\begin{subequations} \label{eq:n2terms}
\begin{align} 
\theta_0 &= 0, \\
q_0 &= q_s, \\
\theta_1 &= - \beta q_s^2 \dd{q_s}{w},  \label{eq:t1defn} \\
q_1 &= q_s \left(\im \theta_1 - \hat{\H}[\theta_1] \right), \label{eq:q1defn}
\end{align}
\end{subequations}
where $\hat{\H} [\theta_1]$ may be evaluated explicitly, so that all terms in \eqref{eq:n2terms} may be written in closed form. Using the relationship \eqref{eq:nohilb}, we may re-write \eqref{eq:compbern} as an equation for $\bar{q}$, the $N=2$ reduced model. This last substitution involves a fair amount of algebra, and so the derivation of this equation is outlined in the appendix. The $N = 2$ reduced model is given by
\begin{align}
& \im \beta \tau \ep^2 \left(q_s + \ep q_1\right) \bar{q}'' + \left( \beta \ep q_s^2 + 2 \beta \ep^2 q_s q_1 - \im \beta \tau \ep^2 q_s' \right) \bar{q}' \nonumber \\
& + \left(- \frac{\im}{q_s} + \ep \frac{\im q_1}{q_s^2} + 2 \beta \ep q_s q_s' \right) \bar{q} + \ep^2 \left( \frac{\im q_1^2}{2q_s^2} + 2 \beta q_s q_s' q_1 + \beta q_s^2 q_1' \right) = 0. \label{eq:N2eqn}
\end{align}
In deriving \eqref{eq:N2eqn} we have chosen to include terms only up to $\Oh(\ep)$, bearing in mind the ansatz \eqref{eq:qexpansatz} which means that derivatives of $\bar{q}$ contribute a factor of $1/\ep$. The leading-order forcing term is $\Oh(\ep^2)$. The reduced model \eqref{eq:N2eqn} is a linear ODE with known coefficients, to be solved for $\bar{q}$. Thus \eqref{eq:N2eqn} is computationally much more convenient than the full system \eqref{eq:gcwave_re},  as the latter includes a non-local term via the Hilbert transform. Moreover, it is much simpler to apply the radiation condition to the reduced model than to the full system, and so we can be sure of selecting the unique physically-relevant solution.

\begin{figure}
\centering
\includegraphics{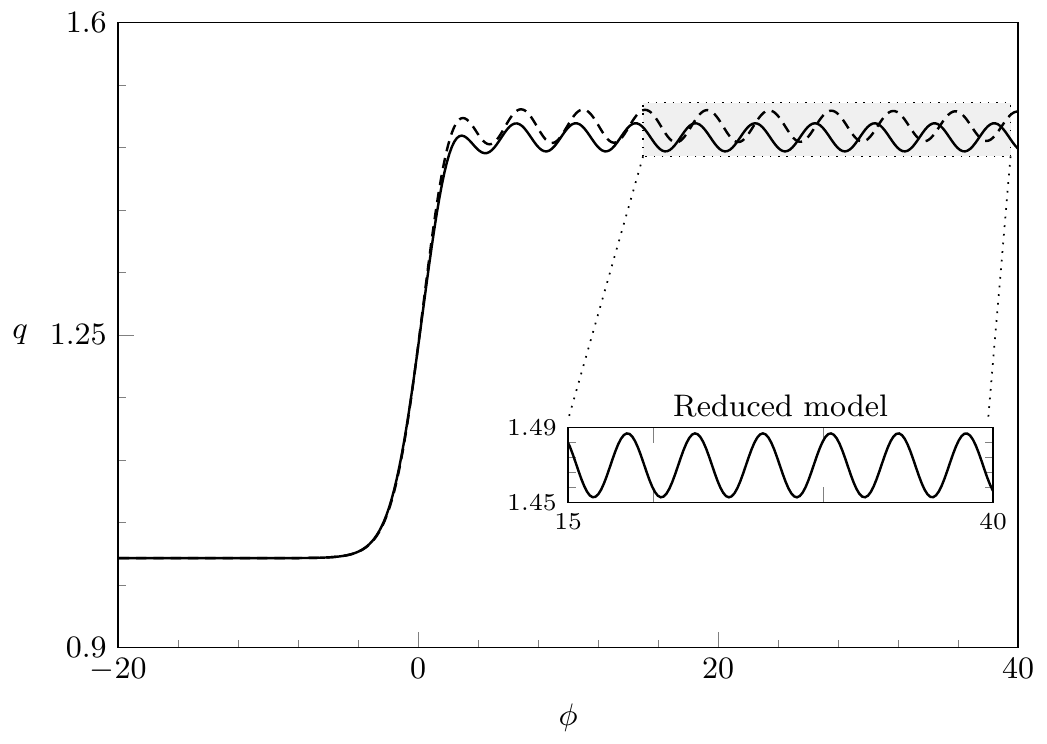}
\caption{Comparison of the full boundary-integral equations \eqref{eq:gcwave_re} (dashed curve) and the $N=2$ reduced model \eqref{eq:N2eqn} (solid curve) for gravity-only flow. The step-height $b = 2$ and $\ep = 0.2$ ($\beta$ may be set to unity without loss of generality). The insets show that the reduced model produces constant-amplitude waves downstream. \label{fig:BDN2comp}}
\end{figure}

Figure \ref{fig:BDN2comp} compares numerical solutions to the full system \eqref{eq:gcwave_re} (dashed curve) and the reduced model \eqref{eq:N2eqn} (solid curve) for gravity-only flow (\emph{i.e.} with $\tau = 0$). In this case, the radiation condition requires that the upstream surface is flat and it is sufficient to solve \eqref{eq:N2eqn} by imposing the condition $\bar{q} = 0$ at the first mesh point (only one condition is needed as the gravity-only problem is first-order). Note that the reduced model produces constant-amplitude waves downstream, whereas the boundary-integral formation does not [cf. figure \ref{fig:stepforbes}(a)]. The reduced model therefore produces constant-amplitude waves of the correct frequency, but nevertheless does not agree with the boundary-integral solution downstream. This is due to truncation at $N=2$ terms.

\subsection{Choosing the boundary conditions} \label{sec:bc}
For gravity-capillary flows, it is insufficient to treat \eqref{eq:N2eqn} as an initial-value problem (\emph{i.e.} `shooting' from upstream). In particular, Type \ta solutions have constant-amplitude capillary waves that extend upstream to $- \infty$, so the values of $\bar{q}$ and $\bar{q}'$ at the first mesh point are not known \emph{a priori}. Even in regimes where the upstream waves decay, however, we found that enforcing a flat surface at the first mesh point leads to inaccurate and unphysical results. Instead, we treat \eqref{eq:N2eqn} as a boundary-value problem with Sommerfeld boundary conditions at either end, using wavenumbers that satisfy the radiation condition. These wavenumbers were derived in section \ref{sec:smallep}. Thus we solve \eqref{eq:N2eqn} with the boundary conditions
\begin{equation} \label{eq:robincond}
\bar{q}' + \im k \bar{q} = 0
\end{equation}
at the first and last mesh points, where $k$ is the relevant wavenumber given by \eqref{eq:kup} and \eqref{eq:kdown} upstream and downstream respectively. By imposing \eqref{eq:robincond} we have restricted ourselves to flows where the far-field solution is a train of linear waves (either constant-amplitude or decaying). However this is clearly not the only possibility, and indeed \eqref{eq:robincond} only allows us to recover one of the four new classes of waves described in \cite{Trinh2013}.

\section{Results} \label{sec:results}

\begin{figure}
\includegraphics{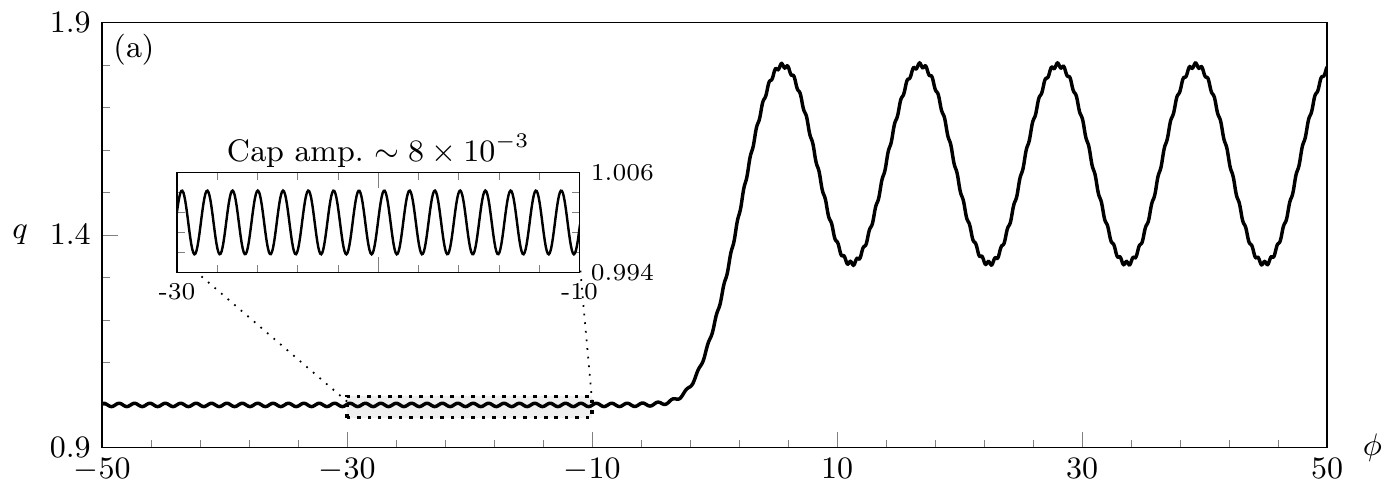} \\
\hfill \\
\includegraphics{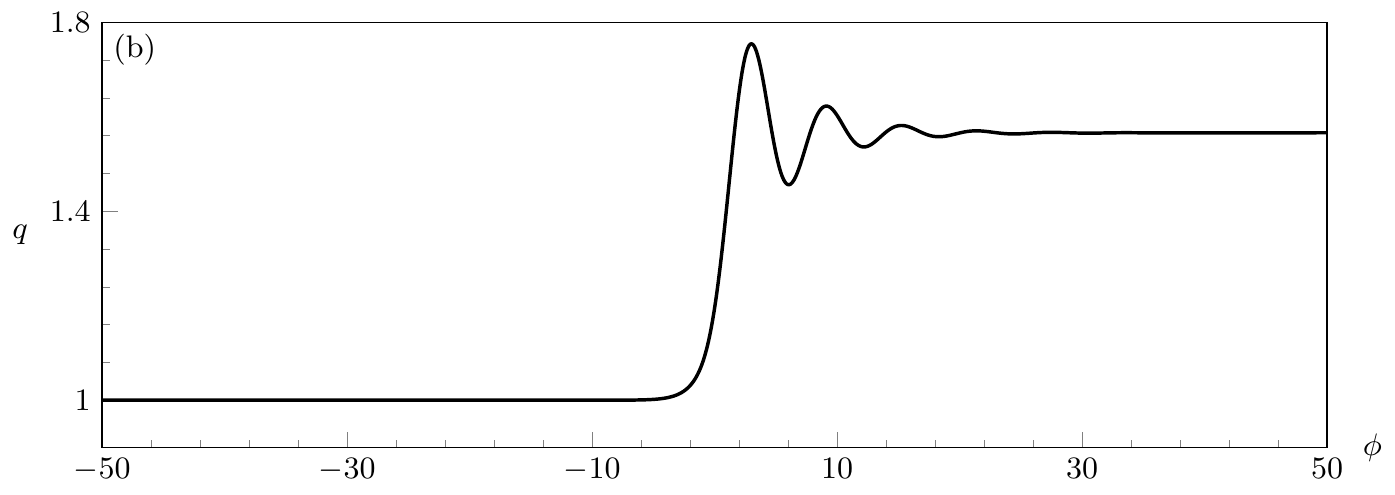}
\caption{Solutions to the $N=2$ reduced model \eqref{eq:N2eqn} as a boundary-value problem (BVP) with Sommerfeld conditions \eqref{eq:robincond} at either end. The parameters are $\{ b, \ep, \beta \} = \{ 2, 0.5, 1 \}$ and the value of $\tau$ is chosen so that each subfigure shows a solution of a different type (see figure \ref{fig:FTbif}). (a) Type \ta solution with $\tau = 0.24$, (b) Type \tb solution with $\tau = 1.5$. Fourier spectra for solution (a) are shown in figure \ref{fig:spectra}(a)-(b). \label{fig:BVPgc}}
\end{figure}

Here we present numerical results for the $N=2$ reduced model, and explore the low-Froude, low-Bond limit of the bifurcation diagram in figure \ref{fig:FTbif} to validate the existence of the new (Type \tc) solutions described in \cite{Trinh2013}. With the boundary conditions \eqref{eq:robincond} applied at either end of the computational domain, we are able to obtain good solutions to the $N=2$ model for a wide range of parameters. By keeping the values of $\beta$, $b$ and $\ep$ fixed (at 1, 2 and 0.5 respectively) and increasing $\tau$ we move up the vertical axis on figure \ref{fig:FTbif} and vary the value of $A = 4\tau/\beta$. 

Figure \ref{fig:BVPgc} shows the two classical solutions that exist when the geometry is linear, but computed with $b = 2$ (\emph{i.e.} in a nonlinear geometry). Figure \ref{fig:BVPgc}(a) shows a type \ta solution with $\tau = 0.24$ and thus $A = 0.96 < 1$, with constant-amplitude waves upstream and downstream. We see that the downstream and, via the inset, upstream solutions are both correctly resolved. The Fourier spectra for this solution are given in figure \ref{fig:spectra}(a, b), and confirm that the dominant wavenumbers agree with the predictions from section \ref{sec:smallep}, denoted G and C, and thus that the radiation condition is satisfied. However, the downstream solution is not completely `clean' and small disturbances are visible on top of the main gravity waves, visible as a small peak in the spectrum at $k \approx 12$; this is shown with an arrow in figure \ref{fig:spectra}(b) and is discussed further in section \ref{sec:beating}. Figure \ref{fig:BVPgc}(b) shows a Type \tb solution, with $\tau = 1.5$ and thus $A = 6 > b^2$. This is the solitary-wave solution from Rayleigh's original classification, and has waves that decay in the far-field both upstream and downstream. 

\begin{figure}
\includegraphics{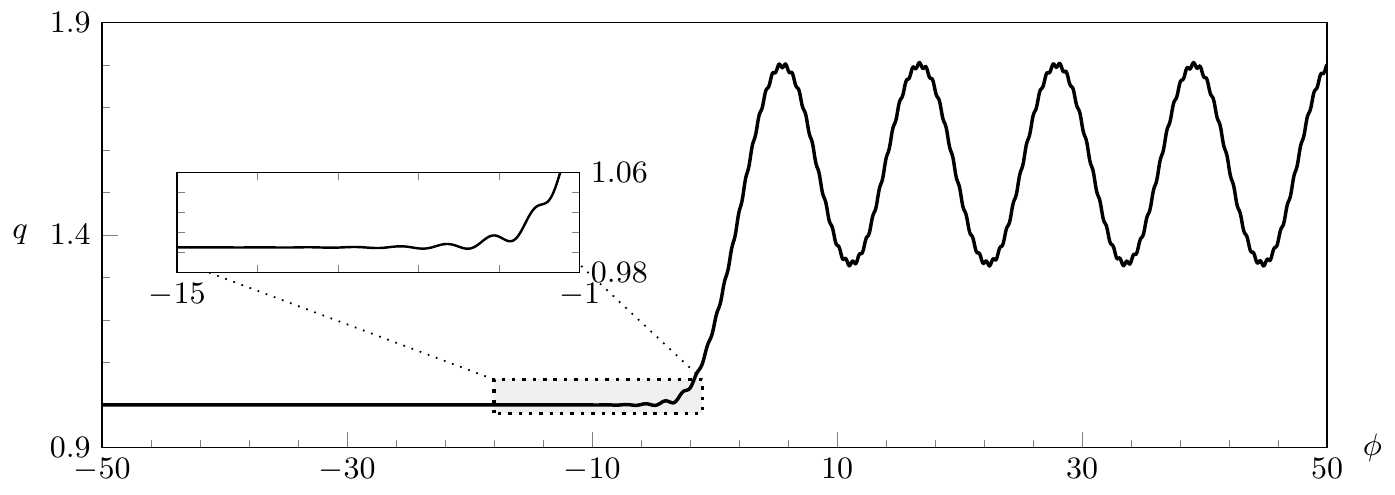}
\caption{Type \tc solution to the $N=2$ reduced model, representing one of the new classes of waves from \cite{Trinh2013}. The parameters are as in figure \ref{fig:BVPgc}, but with $\tau = 0.255$ so that $1 < A < b^2$. The Fourier spectrum of the downstream solution is shown in figure \ref{fig:spectra}(c). \label{fig:BVPnewsol}}
\end{figure}

Figure \ref{fig:BVPnewsol} shows a Type \tc solution with $\tau = 0.255$ and thus $A = 1.02 \in [1, b^2]$. This is one of the new waves from \cite{Trinh2013}, predicted to exist when the step height, $b - 1$, is $\Oh(1)$ in size. There are constant-amplitude waves downstream, but upstream the waves decay away from the step. This is shown in greater detail in the inset. The region over which the waves decay is predicted to increase as $A \to 1$, and tests with various values of $\tau$ (not shown) confirm this. The downstream spectrum is shown in figure~\ref{fig:spectra}(c); and shows that the dominant frequency of the downstream solution is correct, again with a small secondary peak near the label C$^*$.

In summary, the careful application of the Sommerfeld boundary condition \eqref{eq:robincond} to the $N=2$ reduced model \eqref{eq:N2eqn} confirms the existence of one new class of waves predicted in \cite{Trinh2013} for flow over a rectangular step. These solutions satisfy the radiation condition via the Sommerfeld boundary condition, which uses wavenumbers derived in section \ref{sec:smallep}. This is confirmed by the Fourier spectra in figure~\ref{fig:spectra}. However, when $\ep = \Oh(1)$ these wavenumbers are no longer valid, and this leads to interference in the solutions. We will now discuss this in more detail. 

\subsection{The appearance of beating for larger values of $\ep$} \label{sec:beating}

Both the dispersion relation \eqref{eq:smallepdisp} and the $N=2$ reduced model \eqref{eq:N2eqn} are derived from the full water-wave problem in the limit $\ep \to 0$. Thus for $\ep$ sufficiently small, solutions to \eqref{eq:N2eqn} computed using \eqref{eq:smallepdisp} satisfy the radiation condition. For larger values of $\ep$, spurious secondary waves appear in the $N=2$ model, creating the `beating' effect seen downstream in figure~\ref{fig:BVPgc}(a). The secondary waves arise due to interference between the two linearly independent solutions of the second-order ODE \eqref{eq:N2eqn}, which diverges from solutions of the full problem when $\ep = \Oh(1)$. The beating is an indication that the reduced model is being used outside its range of validity, as the low-speed dispersion relation \eqref{eq:smallepdisp} no longer predicts the far-field behaviour of the solution.

\begin{figure}
\centering
\includegraphics{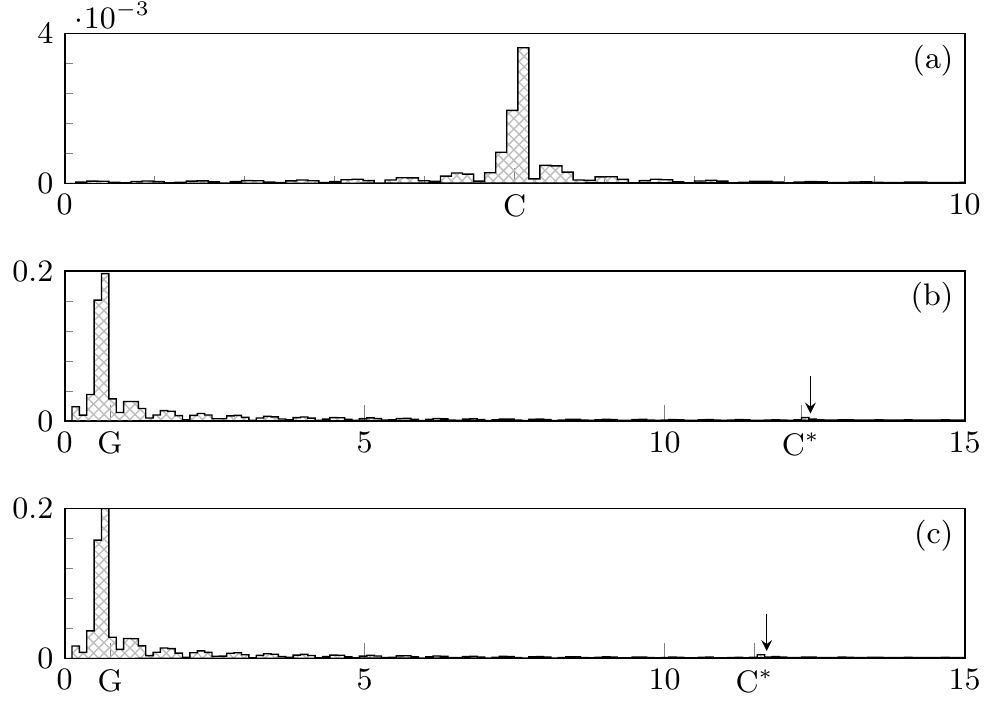}
\caption{(a) Upstream and (b) downstream Fourier spectra for figure \ref{fig:BVPgc}(a). (c) Downstream Fourier spectrum for figure \ref{fig:BVPnewsol}. The horizontal axes correspond to wavenumber $k$ and the vertical axis indicates the absolute value of the $k$-th mode. Labels G and C correspond to gravity and capillary wavenumbers from section \ref{sec:smallep}, while C$^*$ corresponds to the wavenumber from section \ref{sec:beating}.  \label{fig:spectra}}
\end{figure}

To confirm this, we can derive the far-field behaviour of \eqref{eq:N2eqn} directly from the equation itself, and see that it differs from \eqref{eq:smallepdisp} when $\ep = \Oh(1)$. If we let $|\phi| \to \infty$, both $q_0 = q_s$ and $q_1$ tend to constant values:
\begin{align} \label{eq:limits}
q_s \to 1 \quad &\text{and} \quad q_1 \to 0 \quad &\text{as} \quad &\phi \to -\infty, \nonumber \\
q_s \to \sqrt{b} \quad &\text{and} \quad q_1 \to \beta \left( \frac{b^2 - \sqrt{b}}{3 \pi} \right) \quad &\text{as} \quad &\phi \to \infty,
\end{align}
so that far away from the step, \eqref{eq:N2eqn} is a linear ODE with constant coefficients. The two linearly independent solutions have wavenumbers given by 
\begin{subequations} \label{eq:lambdatogether}
\begin{equation} \label{eq:lambda}
\mathcal{K}_{\pm} = \im \frac{ q_s^3 \beta + 2 q_1 q_s^2 \beta \ep \pm \sqrt{\Delta}}{2 q_s \beta \ep (q_s + \ep q_1) \tau}, 
\end{equation}
where
\begin{equation} \label{eq:deltadef}
\Delta = \beta \left( q_s^4 \beta \left(q_s + 2 \ep q_1 \right)^2 - 4(q_s - \ep q_1)(q_s + \ep q_1) \tau \right).
\end{equation}
\end{subequations}
Comparing \eqref{eq:lambda} with \eqref{eq:smallepdisp}, we see that $\mathcal{K}_{\pm}$ agrees with the upstream wavenumber $\ku$ for all $\ep$, but only agrees with the downstream wavenumber $\kd$ in the limit $\ep \to 0$. Thus if $\ku$ and $\kd$ are used in the Sommerfeld boundary conditions when $\ep = \Oh(1)$, beating is observed due to the discrepancy between $\mathcal{K}_{\pm}$ and $\kd$. The downstream spectra shown in figure \ref{fig:spectra}(b)-(c) confirm that the secondary waves visible in figures \ref{fig:BVPgc}(a) and \ref{fig:BVPnewsol} have the frequency predicted by \eqref{eq:lambda} (marked C* on the axis). Thus the interference visible in the solution is due to the use of $\kd$ in the downstream boundary condition in a regime where the far-field solution has frequency $\mathcal{K}_+$. 

Thus the appearance of beating is a sign that the $N = 2$ equation is reaching its limit of validity. Recall that both the dispersion relation \eqref{eq:smallepdisp} and the $N = 2$ reduced model are valid asymptotic limits of the full problem when $\ep \to 0$; however, the wavenumbers $\mathcal{K}_{\pm}$ given in \eqref{eq:lambda} are valid for the $N = 2$ equation at any $\ep$. Thus if the reduced model is to remain relevant in the context of the full problem, it is important that $\ep$ is kept small enough that \eqref{eq:smallepdisp} is still valid and beating does not occur.

\section{Conclusions} \label{sec:conc}
In order to compute state-state gravity-capillary flow past nonlinear geometries, specialised methods are needed to select the solution that satisfies the radiation condition. In section \ref{sec:reduced}, we showed that the full problem can be reduced to a linear ODE using techniques from exponential asymptotics to justify removing the problematic Hilbert transform term. This reduction is valid in the low-Froude, low-Bond limit, but places no restriction on the size of the step. This ODE was still found to be very sensitive in the sense that precise boundary conditions were needed to select solutions that satisfy the radiation condition, but this is computationally simpler to implement in the reduced model than in the boundary-integral framework of the full problem. The relevant boundary conditions were derived in section \ref{sec:disprel} by considering different dispersion relations up- and downstream.

With these boundary conditions, we were able to verify numerically the existence of new waves from \cite{Trinh2013} for flow over a right-angled step, and to confirm that Rayleigh's original bifurcation curve separates into a band when the size of the obstacle is too large to be linearised about. This was done in section \ref{sec:results}, where one of the new classes of waves (decaying upstream, constant-amplitude downstream) was found to exist in an intermediate region between the two classical solutions [cf. figure \ref{fig:BVPnewsol}]. 

\subsection{Difficulties in applications to the full problem}
One of the aims of developing these reduced models is to see whether their solutions can provide a starting guess for the iterative methods needed to solve the system of equations that comprises the full boundary integral problem \eqref{eq:complexgov}. 

The authors ran several numerical experiments, using solutions from the $N=2$ equation \eqref{eq:N2eqn} as starting guesses for the full problem. Experiments were run with a large number ($\approx{2000}$) of grid-points and moderate parameter values ($b = 1.7, \ep = 0.5$) but these failed to converge to the correct solutions. Even on occasions where Newton's method produced a result, the upstream wavenumbers were in serious error, and the accuracy is not significantly improved compared to solutions obtained using the initial guess $\theta = 0$. As discussed in section \ref{sec:truncate}, the $N=2$ model produces solutions that differ from the full problem by a multiplicative constant. This can be seen in the gravity-only case in figure \ref{fig:BDN2comp}, which compares solutions of the $N=2$ model and the full problem. Although the reduced model captures the wavelength correctly, there is a persistent error in amplitude. It is therefore not guaranteed that the two solutions would converge under Newton's method. We do not believe that this conclusion should be regarded as disappointing \emph{per se}, but rather that it illustrates an intrinsic difficulty that has not been noted before in an equally precise fashion.

The sensitivity of the reduced model highlights both the importance and the difficulty of numerically satisfying the radiation condition, and the difficulties in dealing with the full boundary integral problem provide further evidence that the asymptotic approach developed in \cite{Trinh2017reduced} and applied here is a useful one for the study of gravity-capillary flows. We will conclude with a brief overview of different approaches to the radiation problem.

\subsection{A general view of the radiation problem}
In regards to the general radiation problem, we believe there are two possible responses. The first is to design schemes that profit from some asymptotic or analytical property of the solution in order to impose effective boundary conditions on the numerical solver. This is what we have done here; first in simplifying the governing equations to a form that is asymptotically valid at low speeds, and then application of known properties derived using exponential asymptotics. The scheme of \emph{e.g.} \cite{grandison2006truncation}, where the upstream solution is matched to a truncated Fourier series with unknown coefficients is also of this spirit. Our work here has clarified to what extent such asymptotically applied radiation conditions will work and to what extent are they disrupted [cf. section \ref{sec:beating}]. 

In practice, researchers may instead choose to add additional effects (physical or phenomenological) in order to regularize the boundary conditions. For instance, one can solve the time-dependent Euler equations and numerically investigate the $t\to\infty$ limit as in \cite{puaruau2010time}. Or one can apply small artificial damping, $\mu$, so as to destroy the capillary waves upstream as was originally done in \cite{rayleigh1883form}. These procedures may be effective in answering questions regarding the physical phenomena, but it may be extremely difficult to recover certain mathematical structures of the full steady-state problem. For instance, time-dependent formulations will only recover stable configurations (unless, perhaps, time is reversed), or viscous formulations may alter the structure of solutions in an irreparable way. Simply said, it is unclear to what extent the limits $t \to \infty$ or $\mu \to 0$ are distinguished, and whether they account for singular effects in the entire space of steady-state solutions. 
 
\begin{appendix}
\section{Deriving the $N=2$ reduced model} \label{sec:n2derive}
Here, we derive the $N=2$ reduced model given by \eqref{eq:N2eqn}. We first outline the argument from \cite{Trinh2017reduced}, which justifies the removal of the Hilbert transform for the case of gravity-only flow. Then we will describe the algebra that leads to the reduced model for gravity-capillary flow, and include a Mathematica routine that allows for the derivation of higher-order reduced models. 

\subsection{Removing the Hilbert transform: outline}
The analysis required to justify the removal of the Hilbert transform can be easier performed on the gravity-only problem where the equations are first order. Thus we set $\tau = 0$ in the complex governing equations \eqref{eq:complexgov}, and take $\beta=1$ without loss of generality. The dependent variables $q(w)$ and $\theta(w)$ are then split into an $N$-term base series and a remainder as in \eqref{eq:splitting}. The governing equations are then
\begin{align} \label{app:gov1}
\ep \left(q_r^2 q_r' + 2 q_r q_r' \bar{q} + q_r^2 \bar{q}' \right) + \sin{\theta_r} + \bar{\theta} \cos{\theta_r} &= \Oh( \bar{q}^2, \bar{\theta}^2), \\
\log{q_r} - \log{q_s} + \frac{\bar{q}}{q_r} + \hat{\H}[\theta_r] + \hat{\H}[\bar{\theta}] - \im \left( \theta_r  + \bar{\theta}\right) &= \Oh(\bar{q}^2).
\end{align}
These may be combined to give the integro-differential equation
\begin{equation} \label{app:gov2}
\ep \left(q_r' + 2 \frac{q_r' \bar{q}}{q_r} + \bar{q}' \right) + \frac{\sin \theta_r}{q_r^2} - \frac{\cos{\theta_r}}{q_r^2} \left( \im \log{\frac{q_r}{q_s}}  + \im \frac{\bar{q}}{q_r} + \theta_r + \im \hat{\H}[\theta_r] + \im \hat{\H}[\bar{\theta}] \right) = \Oh(\bar{q}^2, \bar{\theta}^2).
\end{equation}
Expanding coefficients in powers of $\ep$, and using the fact that $\theta_0 = 0$ and $q_0 = q_s$, equation \eqref{app:gov2} becomes [cf. (7.2a) in \cite{Trinh2017reduced}]

\begin{equation} \label{app:govfinal}
\ep \bar{q}' + \left[ - \frac{\im}{q_s^3} + \ep \left( 2\frac{q_s'}{q_s} + 3 \im \frac{q_1}{q_s^4} \right) + \Oh(\ep^2) \right] \bar{q} = R(w; \hat{\H}[\bar{\theta}]) + \Oh(\bar{\theta}^2, \bar{q}^2).
\end{equation}
The forcing term $R$ may be written
\begin{equation} \label{Rdef}
R(w; \hat{\H}[\bar{\theta}]) = -\mathcal{E}_{\text{bern}} + \mathcal{E}_{\text{int}} \frac{\cos{\theta_r}}{q_r^2},
\end{equation}
where $\mathcal{E}_{\text{bern}}$ and $\mathcal{E}_{\text{int}}$ are the remainders after the complex governing equations \eqref{eq:complexgov} have been satisfied to $N$ terms by the base series. Note that this corrects equation (7.2) in \cite{Trinh2017reduced}, where $\mathcal{E}_{\text{bern}}$ was multiplied by $\cos{\theta_r}/q_r^2$.

Equation \eqref{app:govfinal} can be solved using the method of integrating factors, giving
\begin{equation}
\ep \bar{q}(w) = Q(w) I(w) \e^{-\chi(w)/\ep} \label{qbarform},
\end{equation}
where
\begin{subequations}
\begin{align}
Q(w) &= \text{const.} \times \frac{1}{q_s(w)^2} \text{exp} \left( -3 \im \int_{w^*}^w \frac{q_1 (t)}{q_s^4 (t)} \mathrm{d}t \right) \label{Qdef} \\
\chi(w) &= - \im \int_{w_0}^w \frac{\mathrm{d}t}{q_s^3 (t)} \label{chidef} \\
I(w) &= \int_{-\infty}^w R(t; \hat{\H}[\bar{\theta}]) \left(\frac{1}{Q(t)} + \Oh(\ep) \right) \e^{\chi(t)/\ep} \mathrm{d}t. \label{Idef}
\end{align}
\end{subequations}
Here $w_0$ is a singularity for $q_s$ in the complex plane and $w^*$ is an arbitrary starting point for the integration. 

As written, \eqref{qbarform} is not a closed expression for $\bar{q}$ since it involves taking the Hilbert transform of the unknown $\bar{\theta}$ in $I(w)$. Note, however, that the exponent $\chi$, which governs the phase of the waves, does not depend on the Hilbert transform at all. The same is true for gravity-capillary flow. As we outline below, $\hat{\H}$ is asymptotically sub-dominant in \eqref{Idef} and so the problem can be vastly simplified.

The function \eqref{Idef} can be analysed using the method of steepest descent (see \cite{trinh2016topological} for more details) in the limit as $\ep \to 0$, where the integral is dominated by the contribution from the endpoints and the singularity $w_0$. Expansion of the integrand at the endpoints shows that the contributions here only serve to provide additional algebraic terms to the expansion and, as is much easier seen through a normal asymptotic expansion, these depend on taking the Hilbert transform of lower-order terms. Thus $\hat{\H}[\theta_r]$ is crucial to determining the later terms of the base series $q_r$. However, when the contribution from the singularity is analysed, we note that 
\begin{equation}
\hat{\H}[\bar{\theta}](w_0) = \frac{1}{\pi} \int_0^{\infty} \frac{\bar{\theta}(\xi')}{\xi' - \e^{-w_0}} \mathrm{d}\xi'
\end{equation}
is small compared to the other terms in \eqref{Rdef}. This is because the singularity $w_0$ lies off the free surface, and so is complex valued. The denominator of the integrand is thus bounded away from zero, and further the remainder $\bar{\theta}$ is expected to be $\Oh(\ep^N)$ on the free surface, where it is evaluated during the integration. Thus, in the steepest descent analysis, the singularity produces the waves without depending on $\hat{\H}[\bar{\theta}]$ at leading order and so these terms can be justifiably removed from the equations without compromising the exponent or the structure of the waves.

\subsection{Derivation of the $N=2$ model}

The reduced model used for the numerical results in section \ref{sec:results} is obtained by truncating the base series after $N=2$ terms. Using the substitution \eqref{eq:splitting}, to leading-order in $\ep$ the complex governing equations \eqref{eq:complexgov} become
\begin{align} \label{eq:o1terms}
\sin{(\theta_0)} &= 0, \nonumber \\
\log{q_0} - \im \theta_0 + \hat{\H}[\theta_0] - \log{q_s} &=0,
\end{align}
whence $\theta_0 = 0$ and $q_0 = q_s$. At the next order,
\begin{align} \label{eq:oepterms}
\beta q_s^2 q_s' + \theta_1 &= 0, \nonumber \\
\frac{q_1}{q_s} - \im \theta_1 + \hat{\H}[\theta_1] & =0,
\end{align}
which gives $\theta_1$ and $q_1$ as in \eqref{eq:t1defn} and \eqref{eq:q1defn}. 

To derive the reduced model, we then substitute 
\begin{equation*}
\theta = - \im \left( \log{\left(q/q_s\right)} + \hat{\H}[\theta] \right)
\end{equation*}
into \eqref{eq:compbern} and expand in powers of $\ep$, with the base series known. By construction, the leading-order forcing term is $\Oh(\ep^2)$. The coefficients of $\bar{q}$ and its derivatives are also expanded in powers of $\ep$, and the number of terms retained reflects the ansatz \eqref{eq:qexpansatz} which indicates that every derivative of $\bar{q}$ contributes a factor of $1/\ep$. For ease of use, we show in Table~\ref{codelisting} how the reduced model can be derived in the coding language Mathematica.

\begin{table}
\begin{lstlisting}
t[w_] = (1/I)(Log[q[w]/qs[w]] + ep Ht[w]);
bern = Sin[t[w]] + beta ep q[w]^2 q'[w] 
        - beta ep^2 tau(q[w] q'[w] t'[w] + q[w]^2 t''[w]);
myeq = bern /. {q -> Function[w, qs[w] + ep q1[w] + ep^2 qb[w]]};
rhs = myeq /. qb -> Function[w, 0];
lhs = myeq - rhs;
n2rhs = Series[rhs, {ep, 0, 2}];
coeffqpp = Series[Coefficient[Series[lhs, {ep,0,5}], qb''[w]], {ep,0,5}];
coeffqp = Series[Coefficient[Series[lhs, {ep,0,5}], qb'[w]], {ep,0,4}];
coeffq = Series[Coefficient[Series[lhs, {ep,0,5}], qb[w]], {ep,0,3}];
\end{lstlisting}
\caption{Mathematica code to generate the reduced model. Functions are written \texttt{q} for $q$, \texttt{qb} for $\bar{q}$ and \texttt{t} for $\theta$. Parameters are written \texttt{ep} for $\ep$, \texttt{tau} for $\tau$ and \texttt{beta} for $\beta$. The Hilbert transform is represented by \texttt{Ht[w]} and to ensure that the asymptotics are done correctly we make the size of the remainder term explicit by writing $\ep^2 \bar{q}$ in place of $\bar{q}$. \label{codelisting}}
\end{table}

\end{appendix}


\end{document}